# EXPLORING THE USE OF CHATGPT BY COMPUTER SCIENCE STUDENTS IN SOFTWARE DEVELOPMENT: APPLICATIONS, ETHICAL CONSIDERATIONS, AND INSIGHTS FOR ENGINEERING EDUCATION

**D. Xu [a], D. Martin [b, 1]**

[a] University College London, London, United Kingdom, 0009-0001-9211-7876
[b] University College London, London, United Kingdom, 0000-0002-9368-4100

## ABSTRACT

ChatGPT has been increasingly used in engineering education, particularly in computer science, offering efficient support across software development tasks. While it helps students navigate programming challenges, its use also raises concerns about academic integrity and overreliance. Despite growing interest in this topic, prior research has largely relied on surveys, emphasizing trends over in-depth analysis of students' strategies and ethical awareness. This study complements existing work through a qualitative investigation of how computer science students in one UK institution strategically and ethically engage with ChatGPT in software development projects. Drawing on semi-structured interviews, it explores two key questions: (1) How do computer science students ethically and strategically report using ChatGPT in software development projects? (2) How do students understand and perceive the ethical issues associated with using ChatGPT in academic and professional contexts? Findings reveal a shift in students' learning models, moving from traditional "independent thinking–manual coding–iterative debugging" to "AI-assisted ideation–interactive programming–collaborative optimization." Importantly, many use ChatGPT conversationally to deepen understanding, while consciously reserving creative and high-level decision-making tasks for themselves. Students tend to cap ChatGPT's contribution to roughly 30%, and evaluate its output to mitigate overreliance. However, only a minority thoroughly analyze AI-generated code, raising concerns about reduced critical engagement. Meanwhile, students reject uncredited use, highlight risks such as privacy breaches and skill degradation, and call for clear usage guidelines set by their teachers. This research offers novel insights into the evolving learner-AI dynamic and highlights the need for explicit guidance to support responsible and pedagogically sound use of such tools.

---

[1] *Corresponding Author*

# 1    INTRODUCTION

Generative Artificial Intelligence (GenAI) is rapidly evolving and permeating education, research, and practice, with the potential to fundamentally transform teaching and learning (Adel et al., 2024). ChatGPT, known for its widespread use and advanced technical capabilities, has become a key support tool in engineering education (Salas-Pilco & Yang, 2022). However, while it offers significant learning advantages, it raises ethical concerns (Fergus et al., 2023). This dual nature underscores the need for deeper research.

The study focuses on computer science (CS) students for their unique role as users who are expected to understand the principles behind GenAI technologies. Their technical training allows them to understand the capabilities and limitations of large language models, recognizing them as "statistical correlation machines" rather than systems based on causal reasoning (Zhou et al., 2024). Additionally, CS students have a heightened awareness of ethical issues, arising from their training and engagement with the societal impacts of AI technologies (Rahman & Watanobe, 2023).

Moreover, a core aspect of CS education are software development projects. They involve multiple dimensions including coding and report writing, making them well-suited for GenAI application research (Akbar et al., 2023). Therefore, this study aims to explore the students' usage patterns and ethical considerations of ChatGPT in software development, focusing on addressing the following two research questions:

How do computer science students ethically and strategically report using ChatGPT in software development projects?

How do students understand and perceive the ethical issues associated with using ChatGPT in academic and professional contexts?

# 2    BACKGROUND

## 2.1 Current Research of GenAI in Engineering Education

The development of GenAI is driving innovation in engineering education, bringing unprecedented possibilities for teaching methods and practical approaches (Nikolic et al., 2023). As a powerful tool, ChatGPT enhances learning efficiency through personalized learning support, adaptive learning tasks, and instant feedback (Maity and Deroy, 2024). It demonstrates unique advantages in simplifying the software development process (Fergus et al., 2023). ChatGPT is widely used in various stages of software development, including code generation, debugging, report writing, and generating test cases, offering efficient solutions to students' programming challenges (Adiguzel et al., 2023). By automating repetitive tasks, these tools allow students to focus on higher-level tasks such as algorithm optimization and problem-solving (Rahman & Watanobe, 2023).

However, the use of this tool also brings potential negative effects and ethical challenges. Overreliance on ChatGPT may undermine students' independent thinking abilities and even lead to academic misconduct, such as plagiarism and code fabrication (Liebrenz et al., 2023). Meanwhile, data privacy, bias, inaccuracies, and overreliance on these tools are recurring themes in current research (Sallam, 2023). These challenges not only affect students' learning processes but also have far-reaching implications for educational equity.

Current research primarily relies on surveys, focusing more on trends rather than providing qualitative analyses of students' thought processes (Alshwiah, 2024). It lacks in-depth exploration of students' rationale for using ChatGPT, which hinders educators' understanding of students' actual needs and behaviors. This study aims to further explore this area by focusing on the specific usage patterns and ethical issues of ChatGPT in software development education, as reported by CS students.

## 3 METHODOLOGY

To explore the usage patterns of ChatGPT in software development education and the ethical issues it raises, this study employed a qualitative research approach, primarily collecting data through semi-structured interviews.

This data collection method was chosen as it allows for an in-depth exploration of students' personal experiences and the behavioral logic underlying their use. Compared to surveys, interviews can capture more nuanced emotions and attitudes, particularly when dealing with complex and multifaceted ethical issues (Denscombe, 2009). This method is better suited to uncovering students' authentic thoughts and concerns (Bell & Waters, 2018).

In previous studies, Dai (2024) used practice observation and design journals to record students' behaviors with ChatGPT, while Guillén-Yparrea and Hernández-Rodríguez (2024) utilized Likert-scale surveys to collect data on students' attitudes toward and frequency of GenAI use. However, observation methods may focus primarily on external behaviors, while surveys may provide limited insight into the underlying motivations and thought processes of students regarding GenAI. By adopting an interview-based approach, this study mitigates these limitations, providing richer and more detailed data while allowing for flexible exploration of emerging themes.

The interviews covered the following aspects:

- How students report using ChatGPT to complete software development tasks.
- Students' understanding and attitudes toward the ethical issues potentially involved in using ChatGPT.

The interviews aimed to capture students' experiences, study how they report interacting with generative AI tools in their learning practices, and explore the underlying logic and attitudes driving these behaviors. The research adhered strictly to institutional ethical guidelines to ensure participant privacy, consent, and the option to withdraw from the study [Ethics reference number: 2025-0753-534].

The participants were 6 students majoring in Computer Science during their bachelor's studies, and currently enrolled at a UK university. Their academic background reflected upper-level undergraduate training in computer science. They had completed core modules in data science, artificial intelligence, and software engineering, with a solid foundation in AI principles relevant to GenAI. Participants were recruited through an open call and voluntarily consented after receiving detailed information about the study. To minimise potential power dynamics, neither of the researchers had any teaching or supervisory relationship with the students. Interviewees were selected for their familiarity with ChatGPT in software development and their contextual understanding of engineering education.

The interviews were conducted via online Zoom meetings. Each interview lasted around 30 minutes and covered students' experiences of using ChatGPT, perceived benefits and drawbacks, ethical concerns, and suggestions for future educational applications. Before interviews, participants signed consent forms and were briefed on the study's purpose, data usage, and privacy measures. Interviews were audio-recorded and transcribed.

The interview transcripts were analysed using thematic analysis to identify patterns and ethical dilemmas associated with ChatGPT usage. First, author 1 transcribed and closely read the interviews to develop familiarity and take initial notes. The next stage involved coding and theme extraction. Initial codes were consolidated, and key themes and subthemes were identified by the first author. Finally, the key themes were interpreted and synthesized in relation to the research questions.

To ensure the reliability of findings, the two authors tracked the coding and discussed the assigned codes and the rationale for coding specific excerpts to ensure consensus across the coders' interpretations. A focused IRR approach was adopted using one coder as reference, and calculated conditional agreement. This approach is particularly suitable when the objective is to ensure reliability in identifying critical codes (Campell, 2013). Among the 25 data segments coded by the first author, 23 were also independently coded into the same categories by the second author, yielding a percent agreement rate of 92%, which exceeds the commonly accepted threshold of 80% (Bazen, 2021). Discrepancies between coders were analyzed and discussed, helping to refine code definitions and enhance analytical clarity. Intercoder reliability ensures that the coding process is not idiosyncratic to a single researcher and supports analytical transparency and replicability (Cole, 2024).

Considering the limitations of this study, there is currently no widely accepted standard for evaluating AI contributions, and the study relies on self-reported data, which may introduce subjective bias (Kasneci et al., 2023). For example, students' estimates of AI-generated content (e.g., 30%) are based on personal perception, rather than objective measures such as the time spent using AI tools, the number of AI-generated outputs, or interaction logs. Third, the study lacks multiple data sources for validation, such as observational data or log analysis, which reduces the robustness of the findings and may overlook important patterns. Future research could address this limitation by incorporating multiple data sources for validation and developing objective standards for assessing AI contributions. These improvements would lead to more reliable and comprehensive insights into students' use of GenAI tools.

# 4   RESULTS

Through interviews, this study identifies several key findings centered around the strategic use of ChatGPT, its ethical boundaries, and its implications for AI ethics training in engineering education (Table 1).  While some of these findings align with prior research on students pursuing engineering degrees, computer science students also exhibited unique characteristics in their interactions with ChatGPT.

*Table 1. Key Findings on Student Usage, Perceptions, and Suggestions for ChatGPT*

| Category | Research Findings |
|---|---|
| **Ways of using ChatGPT** | Students use ChatGPT for task analysis, brainstorming, coding, debugging, testing, and report writing. |
| | ChatGPT provoke conversational learning, deepening understanding and provoking ideas. |
| | Non-native speakers improve language expression for reports. |

| | Students prefer handling creative and decision-making tasks independently, using ChatGPT only as a reference tool. |
|---|---|
| **Balancing independent thinking and dependency** | Students limit ChatGPT's contribution (about 30%) and review and adjust its outputs. |
| | Only few study and comprehend AI-generated codes line by line, most reduce critical thinking when outputs seem correct, neglecting in-depth understanding. |
| | Students avoid being limited by ChatGPT by discussing ideas in teams or thinking independently before using it for refinement. |
| **Ethics and risk awareness** | Students strongly oppose using ChatGPT outputs directly, considering it a violation of academic integrity. |
| | Students emphasize acknowledging ChatGPT's contributions for academic transparency. |
| | Concerns include privacy risks, potential copyright issues, and skill degradation from over-reliance. |
| **Suggestions** | Define proper ChatGPT usage, teach effective prompt design, and encourage critical analysis. |
| | Require clear acknowledgment of ChatGPT's involvement. |

## 4.1 Strategic Use of ChatGPT

### 4.1.1 ChatGPT Integration Across the Software Development Lifecycle

Students primarily used ChatGPT for tasks such as clarifying doubts about assignments, brainstorming research ideas, generating code, providing explanations, testing, debugging, and writing reports. This pattern aligns with the findings from other studies (Champa et al., 2024; Akbar et al., 2023).

Students indicated that the first thing they do upon learning about an assignment is to use ChatGPT to help them understand the task requirements and provide relevant background knowledge. As interviewee 4 stated: "I usually start by discussing the assignment requirements from the instructor with the AI tool". Interviewee 1 further explained that it helped to quickly grasp core concepts and historical context in a field, saving significant time.

Students are forming a conversational learning habit with ChatGPT, where interactions go beyond simple answer retrieval and increasingly serve as a cognitive tool that provoked their thinking. Interviewee 3 viewed ChatGPT as a thinking partner, noting that its explanations often offered new insights and deepened their understanding of the problem. Interviewee 5 further stated: "This often inspires my thinking. ChatGPT tends to consider two steps ahead, helping me broaden my problem-solving approach."

Non-native English-speaking students also leveraged ChatGPT for language optimization, improving the grammar of their reports. As interviewee 2 noted, it made their documents look more professional and easier to understand.

Meanwhile, students explained that they prefer to handle certain tasks on their own. This includes creative and decision-making activities such as software architecture design and complex algorithm development, which require deep thinking and originality. Interviewee 2 highlighted the importance of product and architecture design, noting that these tasks demand creativity and thoughtful consideration, with ChatGPT only offering preliminary suggestions. Interviewee 6 explained: "Creative and innovative tasks—such as solving complex interdisciplinary problems or proposing novel scientific hypotheses—require flexible, human-style reasoning and intuition." Students also emphasized the need to complete original work independently, such as taking quizzes, to maintain academic integrity. Interviewee 3 emphasized that academic integrity-related tasks such as original paper writing, reflective writing, and exams must be completed independently by humans.

### 4.1.2 Balancing ChatGPT Assistance and Independent Thinking

The interviews revealed that students were mindful of maintaining boundaries when using ChatGPT, with its contributions to software development projects consistently kept below 30% according to self-reports. Students consciously and consistently maintained critical thinking throughout the process.

Students did not fully trust ChatGPT, believing its generated content requires further thought and refinement, treating it as an initial proposal or reference rather than a final solution. Interviewee 3 noted: "After getting its output, I always think about whether it really fits my project or my needs." Interviewee 4 explained: "A single problem can have many possible approaches—the AI helps suggest them, but it's up to me to choose and refine one." Interviewee 1 summarized: "ChatGPT is more of an assistant; the final code needs to be modified and refined by me."

However, a notable concern in the usage patterns is that only less than half of the interviewees reported carefully analysing and reviewing each line of AI-generated code to ensure a thorough understanding. Most students admitted that using ChatGPT can make it hard to maintain independent thinking. Interviewee 2 noted that if ChatGPT's responses seem reasonable and align with their ideas, they might be influenced by it and stop thinking deeply about certain issues. This reliance could lead to overlooking the process of critical thinking—for example, focusing solely on solving the problem without understanding why the error occurred, and failing to truly internalize the problem-solving process.

At the same time, students also have their own strategies to try to maintain independent thinking. For example, interviewee 5 described writing her own code first, then using ChatGPT to assess its quality and alignment with instructional goals. Interviewee 1 emphasized actively participating in team discussions and sharing ideas to refine their own perspectives and avoid being limited by ChatGPT's viewpoints.

## 4.2 Ethical Boundaries and Potential Risks in ChatGPT Usage

### 4.2.1 Awareness of Ethical Boundaries

Students demonstrated a clear ethical stance on using ChatGPT in academic settings. First, students explicitly opposed the direct use of AI-generated outputs. Interviewee 2 emphasized, "If you directly use ChatGPT's outputs, such as generating a complete PPT or code with a single prompt, that's definitely not appropriate." Second, students stressed the importance of acknowledging ChatGPT's contributions. Interviewee 1 stated, "When presenting results, I clearly explain that ChatGPT provided foundational ideas and reference information, which I further analysed, adjusted, and innovated upon to achieve my research goals." This reflects a commitment to academic integrity and transparency in recognizing external assistance.

### 4.2.2 Potential Risks and Common Concerns

For data privacy issues, students expressed concerns about ChatGPT's cloud-based operation potentially leading to sensitive data leaks. Interviewee 6 stated: "I've also noticed that AI retains previous conversations and may reuse personal details without explicit consent." Students expressed that they avoided inputting personal or confidential data into ChatGPT to prevent potential data leaks.

Regarding code copyright issues, they noted that generated code might resemble existing open or closed-source code, posing risks of copyright infringement（Lucchi, 2024）. Interviewee 6 explained: "Relying entirely on AI would not only violate academic standards but might also unintentionally infringe on others' intellectual property, as AI-generated content is often based on existing data." Students also highlighted the potential for skill degradation, warning that over-reliance on ChatGPT could erode their independent problem-solving and programming skills. As interviewee 2 remarked: "If ChatGPT suddenly went offline, I would really struggle to complete this kind of assignment from scratch."

Finally, students raised concerns about contribution attribution, expressing confusion over how to quantify AI's indirect contributions. Interviewee 3 explained: "The solutions provided by ChatGPT inspire my ideas, but the final implementation is mine. It's hard to evaluate this indirect contribution."

### 4.3 Implications for GenAI Ethics Training in Engineering Education

Students suggested clarifying ChatGPT's role and usage rules in educational settings. They proposed defining tasks that ChatGPT can assist and those that require independent completion. Interviewee 5 pointed out: "Most of my instructors say AI can be used as a support tool, but the definition of 'support' is vague." Interviewee 4 further stated: "I think schools should clearly define the rules for using AI: what is and isn't allowed." Meanwhile, they recommended requiring students to specify ChatGPT's involvement in reports to uphold academic integrity. Furthermore, students suggested teaching how to craft effective prompts and critically analyse ChatGPT's outputs.

## DISCUSSION AND CONCLUSION

The analysis of interviews revealed how students with CS background are strategically and ethically using ChatGPT for software development assignments, leveraging its efficiency for repetitive tasks while maintaining dominance over creative tasks. Students' views offered critical insights that may contribute to constructing a GenAI ethics training framework. This framework should clarify usage boundaries, promote critical and independent thinking, strengthen ethical awareness, and provide technical training to better utilize GenAI tools.

The findings of this study reveal a major transformation in students' learning approaches due to the integration of ChatGPT. The traditional "independent thinking-coding practice-debugging optimization" learning model is gradually evolving into an "AI-assisted thinking-interactive programming-collaborative optimization" model. This transformation aligns with the findings of Rahman and Watanobe (2023), who noted that generative AI is fundamentally reshaping traditional learning trajectories.

Students applied ChatGPT across multiple stages of software development projects. During the requirement analysis stage, ChatGPT was used to supplement information and expand research perspectives. In the code architecture design stage, it provided initial suggestions, while core design work remained a human responsibility. During code implementation, ChatGPT served as a reference for code generation. For code debugging, it analysed errors and suggested possible solutions. Finally, in report writing, ChatGPT was used to enhance language clarity and logical flow.

Additionally, interviewees generally gave higher evaluations of the effectiveness of ChatGPT and reported a higher frequency of ChatGPT usage compared to the average levels observed in prior research (e.g., Dai, 2024). This finding aligns with the perspective that sufficient background knowledge is necessary to benefit from ChatGPT (Daher & Hussein, 2024). Rodrigues et al. (2021) stated that students need knowledge and skills to thrive in the digital age, suggesting that the CS background of our participants may have equipped them with the technical understanding needed to leverage these AI tools more effectively. Notably, all the participants implemented self-regulation strategies to ensure independent thinking when using ChatGPT, consistently keeping its contribution below 30%, aligning with the critical awareness of AI tools described by Farangi et al. (2024).

Interviewed students exhibited ethical awareness. Students demonstrated the ability to clearly distinguish between tasks suitable and unsuitable for AI tools and spontaneously formed ethical guidelines concerning data privacy and academic integrity. This spontaneous ethical consciousness offers a valuable reference point for building more systematic AI ethics education frameworks.

Published research identifies several ethical concerns associated with ChatGPT. These include hallucination, where it generates plausible but incorrect information (Stahl & Eke, 2024); lack of originality, as its outputs are recombinations of training data; and bias, reflecting imbalances in its training sources (Hua, Jin, & Jiang, 2024). Additionally, privacy risks arise from large-scale data use, while sustainability concerns stem from the high financial, environmental, and human resource costs of AI systems (Mannuru et al., 2023). Compared to prior research, students in this study displayed a higher level of critical engagement with ChatGPT's outputs, refusing to accept them uncritically. As CS students, their technical expertise likely contributed to their skepticism, allowing them to recognize the probabilistic nature and reasoning limitations of large language models. While all participants mentioned issues such as hallucination, originality, and privacy risks, there were no mentions to bias and sustainability. This suggests the need for computer science education to expand beyond technical aspects and place greater emphasis on sustainability and macroethical considerations, as suggested by Grosz et al. (2019)

Several implications for engineering education emerge. The study draws on student views and connect it to prior research highlighting the importance of ethical training, clear guidelines, and transparency to uphold academic integrity (Cotton et al., 2023; Atlas, 2023; Cooper, 2023). Universities should provide all students with foundational knowledge of generative AI to enable critical engagement. Targeted workshops on prompt engineering, AI output evaluation, and privacy concerns could help students optimize AI usage while mitigating risks. This training could be complemented by information on sustainability issues purporting to the use of large language models and how to evaluate and interpret outputs as biased or non-biased. Clear academic guidelines defining ChatGPT's role in coursework would provide structured boundaries for ethical use. Establishing categories of allowed, partially allowed, and prohibited AI usage in academic settings, along with transparency requirements for disclosing AI involvement, would enhance clarity and integrity. Additionally, educators could balance AI reliance with skill development by using AI-generated content as a baseline and incorporating reflection reports to encourage critical engagement. Together, these insights point toward a more comprehensive and student-informed framework for supporting ethical and effective use of AI in engineering education.